\documentclass[%
 reprint,
superscriptaddress,
 amsmath,amssymb,
 aps,
 nofootinbib,
floatfix,
]{revtex4-1}

\usepackage{bm}
\usepackage{amsmath,amssymb}
\usepackage{amsfonts}
\usepackage{layouts}
\usepackage{mathtools}
\usepackage{dsfont}
\usepackage{graphicx}
\usepackage{float}
\usepackage{subfigure} 
\usepackage{verbatim}
\usepackage{amsfonts}
\usepackage{bbold}
\usepackage{dcolumn}
\usepackage{bm}
\usepackage{color}
\usepackage[dvipsnames]{xcolor}
\usepackage{listings}
\definecolor{blueprl}{RGB}{46,48,146}

\usepackage{soul}

\usepackage{booktabs}

\usepackage{mathrsfs}
\usepackage{filecontents}
\usepackage{times} 

\newcommand{\ensavg}[1]{\langle#1\rangle}

\newcommand{\xdownarrow}[1]{%
  {\left\downarrow\vbox to #1{}\right.\kern-\nulldelimiterspace}
}

\makeatletter
\newcommand*{\balancecolsandclearpage}{%
  \close@column@grid
  \clearpage
  \twocolumngrid
}
\makeatother

\usepackage{gensymb}
\usepackage[breaklinks=true]{hyperref}
\hypersetup{
colorlinks   = true, 
urlcolor     = blue, 
linkcolor    = blue, 
citecolor    = magenta 
}
\usepackage{comment}
\usepackage[normalem]{ulem}
\usepackage{xcolor}
\usepackage{graphicx}
\usepackage{cleveref}
\crefname{equation}{Eq.}{Eqs.}
\Crefname{equation}{Equation}{Equations}
\crefname{figure}{Fig.}{Figs.}
\Crefname{figure}{Figure}{Figures}
\crefname{figure}{Fig.}{Figs.}
\Crefname{figure}{Figure}{Figures}
\crefname{section}{}{}
\Crefname{section}{}{}
\crefname{appendix}{Appendix}{Appendices}
\Crefname{appendix}{Appendix}{Appendices}
\crefname{table}{Table}{Tables}
\Crefname{table}{Table}{Tables}

\usepackage[most]{tcolorbox}

\tcbset{textmarker/.style={%
        enhanced,
        parbox=false,boxrule=0mm,boxsep=0mm,arc=0mm,
        outer arc=0mm,left=6mm,right=3mm,top=7pt,bottom=7pt,
        toptitle=1mm,bottomtitle=1mm,oversize}}

\newtcolorbox{hintBox}{textmarker,
    borderline west={6pt}{0pt}{yellow},
    colback=yellow!10!white}
\newtcolorbox{noteBox}{textmarker,
borderline west={-1pt}{-1pt}{white},
colback=blue!10!white}
\newtcolorbox{importantBox}{textmarker,
borderline west={-1pt}{-1pt}{white},
colback=red!10!white}

\definecolor{Midnight_Blue}{rgb}{0.1, 0.1, 0.6}

\makeatletter
\def\@bibdataout@aps{%
 \immediate\write\@bibdataout{%
  @CONTROL{%
   apsrev41Control,author="08",editor="1",pages="0",title="0",year="1",eprint="1"%
  }%
 }%
 \if@filesw
  \immediate\write\@auxout{\string\citation{apsrev41Control}}%
 \fi
}%
\makeatother 

\AtBeginDocument{%
  \heavyrulewidth=.08em
  \lightrulewidth=.05em
  \cmidrulewidth=.03em
  \belowrulesep=.65ex
  \belowbottomsep=0pt
  \aboverulesep=.4ex
  \abovetopsep=0pt
  \cmidrulesep=\doublerulesep
  \cmidrulekern=.5em
  \defaultaddspace=.5em
}

\begin{document}


\title{Classical-Quantum Dual Encoding for Laser Communications in Space \\
}

\author{Matthew S. Winnel}
\affiliation{Centre for Quantum Computation and Communication Technology, School of Mathematics and Physics, University of Queensland, St Lucia, Queensland 4072, Australia}
\author{Ziqing Wang}
\affiliation{School of Electrical Engineering \& Telecommunications, The University of New South Wales, Sydney, NSW 2052, Australia}
\author{Robert Malaney}
\affiliation{School of Electrical Engineering \& Telecommunications, The University of New South Wales, Sydney, NSW 2052, Australia}
\author{Ryan Aguinaldo}
\affiliation{Northrop Grumman Corporation, San Diego, California 92128, USA}
\author{Jonathan Green}
\affiliation{Northrop Grumman Corporation, San Diego, California 92128, USA}
\author{Timothy C. Ralph}\email{ralph@physics.uq.edu.au}
\affiliation{Centre for Quantum Computation and Communication Technology, School of Mathematics and Physics, University of Queensland, St Lucia, Queensland 4072, Australia}

\date{\today}

\begin{abstract}
{In typical laser communications classical information is encoded by modulating the amplitude of the laser beam and measured via direct detection. We add a layer of security using quantum physics to this standard scheme, applicable to free-space channels. We consider a simultaneous classical-quantum communication scheme where the classical information is encoded in the usual way and the quantum information is encoded as fluctuations of a sub-Poissonian noise-floor. For secret key generation, we consider a continuous-variable quantum key distribution protocol (CVQKD) using a Gaussian ensemble of squeezed states and direct detection. Under the assumption of passive attacks secure key generation and classical communication can proceed simultaneously. Compared with standard CVQKD. which is secure against unrestricted attacks, our added layer of quantum security is simple to implement, robust and does not affect classical data rates.  We perform detailed simulations of the performance of the protocol for a free-space atmospheric channel. We analyse security of the CVQKD protocol in the composable finite-size regime.}
\end{abstract}

\maketitle

\section{Introduction}

There has been increased interest for modern satellite systems to adopt optical communications, also known as laser communications in space (LCS). The advantages include higher rates, longer distances, increased security because of strong directionality, and lack of frequency license regulations since the radio frequency bands are close to being saturated \cite{TOY20}. In typical laser communications, information is encoded by modulating the amplitude of the laser beam and measured via direct detection.

An additional feature of LCS is that the optical channel is also suitable for quantum communications allowing the possibility to combine classical and quantum communications \cite{TAK17}. Previous work on this topic has started with quantum communication, which is complex and has poor tolerance to noise, and added classical communication \cite{KUM19}. Here, we consider an alternative approach in which we start with a standard classical communication set-up, and then introduce quantum techniques to enhance security.

Specifically, in this work, 
we incorporate quantum key distribution~\cite{RevModPhys.81.1301,Pirandola_2020} with a standard classical communication set-up. Quantum key distribution (QKD)~\cite{RevModPhys.81.1301,Pirandola_2020,Xu_2020,Curty_2004} is the most mature and accessible quantum technology. Its goal is to distribute a secret random key between two physically-separated parties. Its unconditional security relies on quantum physics, a promising solution to the vulnerability of current classical cryptosystems to quantum computing.

Encodings of quantum information can be characterised as discrete variable (DV) and continuous variable (CV)~\cite{Braunstein_2005,yonezawa2008continuousvariable,cerf2007quantum,Weedbrook_2012,Serafini2017QuantumCV}. CV encodings exploit the infinite-dimensional Hilbert space of an oscillator, i.e., they encode information into the CV quadratures of the light, such as amplitude $\hat{q}$ and phase $\hat{p}$ which have a continuous spectrum. CV~\cite{8439931} has the advantage of easier state preparation and manipulation~\cite{Braunstein_2005} as well as being compatible with existing infrastructure~\cite{Kumar_2015}. Compared with standard CV QKD~\cite{PhysRevLett.88.057902,PhysRevLett.93.170504,Gottesman_2001,PhysRevA.63.052311}  our layer of quantum security is easy to implement, has robust performance and is directly compatible with LCS. The trade-off is our protocol is only secure against passive attacks, unlike standard CV QKD which is secure against unrestricted eavesdropping attacks~\cite{PhysRevLett.118.200501,PhysRevLett.109.100502,PhysRevLett.97.190503,PhysRevA.99.012311}.



Satellite-based quantum communication~\cite{Bonato_2009,5342318} promises global-scale quantum networks by being able to connect any two points on Earth. The dominant source of noise for quantum communication comes in the form of losses~\cite{cerf2007quantum}, for instance, scattering and absorption loss in fibres, beam wandering and scattering in atmospheric links, and beam spreading in free space. Unlike fibre links, the direct line-of-sight link can be monitored by the trusted parties, meaning the eavesdropper (Eve) has limited access to the channel. As a result the only plausible attack on an LCS system is a passive one in which lost light is intercepted by the eavesdropper~\cite{https://doi.org/10.48550/arxiv.2212.04807, PhysRevA.105.032602}.

During a point-to-point quantum communication protocol, quantum correlations exist between all three parties; the sender (Alice), the receiver (Bob), and the potential eavesdropper (Eve). Since information is inevitably leaked to Eve, the correlations between Bob and Eve must be suppressed to maintain security; for example, via privacy amplification. This can only work if Alice and Bob's mutual information is greater than Eve's maximal information with Bob~\cite{devetak2005distillation}. Alternatively, Alice and Bob could use entanglement purification~\cite{bennett1996purification,https://doi.org/10.48550/arxiv.2203.13924} or entanglement distillation~\cite{PhysRevA.100.022315} to suppress Eve's information~\cite{Braunstein_2005} in an entanglement-based QKD protocol.

In Ref.~\cite{Jacobsen2018} a different approach was taken whereby they introduced a zero-leakage protocol, with zero information leakage at all times in a pure-loss channel, by encoding quantum information in a particular way. In this protocol the signal is only on one quadrature, although the noise properties of both quadratures are monitored. Ref~\cite{PhysRevA.104.012411} introduced a symmetric version of this protocol whereby information is encoded on both quadratures. However, this is experimentally more complicated. Ref.~\cite{PhysRevA.105.032602} investigated the zero-leakage protocol in the context of satellite-based CV QKD, restricting Eve to passive attacks; 
thus, simplifying generation of a secret key and increasing tolerance to loss and noise, while also eliminating the need for a local oscillator. 

In this paper, we show how a satellite-based zero-leakage quantum communication protocol can be combined with a LCS system without affecting the classical data rates. To be consistent with the zero-leakage protocol we restrict to a passive eavesdropper (Eve), who may collect all the light lost but may not perform any active attack. For example we do not allow Eve to intercept and resend Alice's beam as she could be observed in the line of sight. More sophisticated entangling cloner attacks \cite{GRO03} are not allowed as they also require intervention in the line of sight. We make a detailed evaluation of our dual classical/quantum LCS system in the context of a satellite to ground down link.





The rest of this paper is  organised as follows. In Section~\cref{sec:the_protocol} we specify the protocol in detail and describe its performance under ideal conditions.
In Section~\cref{Section:SimulationSettings} we consider the satellite-to-ground communication set-up and investigate atmospheric effects. In Section ~\cref{sec:secret_key} we combine our protocol description with our atmospheric characterization and compute secret key rates in a realistic finite-size regimes. In Section~\cref{sec:conclusion} we conclude.

\section{The protocol}\label{sec:the_protocol}

\begin{figure}
\centering
\includegraphics[width=1\linewidth]{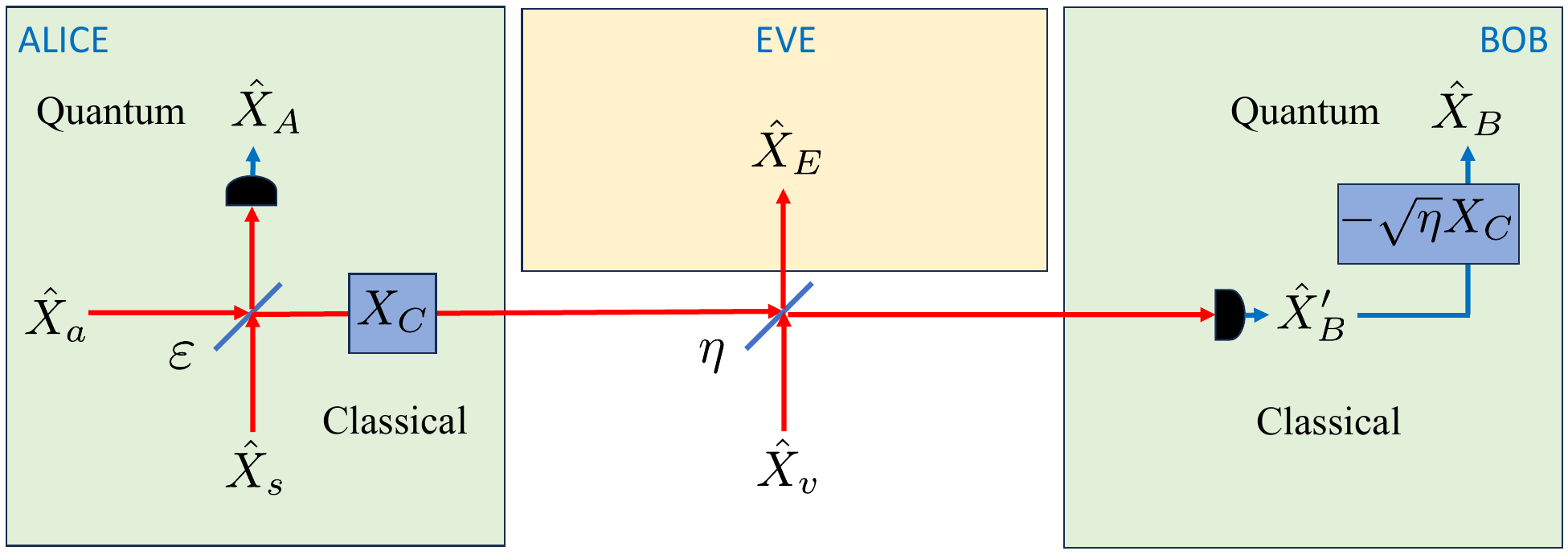}
\caption{Protocol set-up. Alice mixes an amplitude squeezed vacuum state, $X_s$, (with $V_s < 1$) and a bright beam with anti-squeezed amplitude fluctuations, $X_a$, (with $V_a > 1$) on a beamsplitter with transmissivity $\varepsilon$. Classical information, $X_C$,  is then encoded on the transmitted beam as shown. Direct detection of the reflected beam reveals Alice's copy of the quantum signal, $\hat X_A$. The transmitted beam travels through the channel which is monitored by Eve who obtains $\hat X_E$. Bob measures the state after the channel with direct detection to obtain his copy of the classical information, $\hat X'_B$. After determining the classical bit values he can extract his copy of the quantum signal, $\hat X_B$. When the variance of the transmitted ensemble equals the QNL, there is no correlation between $\hat X_E$ and $\hat X_B$ for a pure-loss channel.}\label{fig:set-up}
\end{figure}

The protocol consists of sending classical information and quantum information simultaneously. Both the classical information and quantum information are encoded as amplitude modulations of a bright laser beam. The classical information is encoded as discrete modulations whilst the quantum information is encoded as smaller Gaussian modulations. To optimize capacity we assume the classical and quantum modulations are encoded onto the same frequency side bands. The quantum modulations effectively set the noise floor for the classical modulations. Using squeezing it is arranged that this noise floor is the quantum noise limit (QNL), also known as shot noise. Both quantum and classical signals are read out via direct detection. A diagram of the set-up is shown in~\cref{fig:set-up}. Alice is the sender, Bob is the receiver and Eve is a passive eavesdropper who captures all the lost light.

While any quantum communication encoding can be used to also encode classical information, it is not optimal to let the low amplitude quantum modulations for QKD directly carry the classical signals due to the low signal to noise. Here, we combine using low amplitude quantum signals with the higher amplitude classical signals for efficient transmission of both quantum and classical information.

\subsection{Classical part}

For the purpose of classical optical communications, the Intensity Modulation (IM) technique (which maps a classical information bit to one of the output intensities of a laser beam), together with the Direct Detection (DD) technique (which directly measures the intensity of the received signal using a photodetector), constitutes the widely-used IM-DD scheme. 
Despite its low implementational complexity, the IM-DD scheme is not desirable for our purpose mainly due to the IM technique's incompatibility with the quantum part of our protocol. 

Instead, in this work 
we adopt a discrete modulation scheme where every classical information bit is mapped to one of the two real-valued displacements $\{\alpha,-\alpha\}$ applied to a continuous-wave single-mode laser beam with optical frequency $\Omega_0$ and overall amplitude $\beta >> |\alpha|$. 
This scheme resembles the Binary Phase-Shift Keying (BPSK) modulation scheme since it only encodes information onto one quadrature (i.e., the $X$ quadrature) of the laser beam.
Our modulation at a specific sinusoidal modulation frequency, $\omega$, creates an \textit{upper sideband} and a \textit{lower sideband} at $\Omega_0+\omega$ and $\Omega_0-\omega$, respectively.
The component remaining at $\Omega_0$ is known as the \textit{carrier}, which can be considered as a strong coherent component~\cite{BAC19,COL87}. We will show in the following that although our protocol does not use IM for modulation, it can still benefit from a reduced implementational complexity provided by the use of DD for detection.

Traditionally, the extraction of quadrature-encoded information requires a coherent homodyne detection involving the mixing of the signal with an \textit{independent} Local Oscillator (LO). By independent LO, we mean a bright laser beam that is phase-locked to the signal and typically sent along with the signal via multiplexing. 
In this work, we eliminate the need for such an independent LO by adopting the DD technique, which effectively measures the $X$ quadrature (i.e., the only quadrature used for information encoding in our scheme) using a photodetector, for signal detection.
To see how the DD technique works, we first consider a propagating continuous-wave laser beam in a single mode represented by the continuum operator $\hat{A}$:
\begin{align}
    \hat{A} &= \beta + \delta\hat{A}(t),
\end{align}
where $\beta$ is the mean value of the carrier amplitude (taken to be real) and $\delta\hat{A}(t)$ represents the modulations (including both the discrete classical modulation and the quantum fluctuations) around the mean value. 
The notation $\delta$ emphasises that the modulations are small compared with $\beta$ (note that $\langle \hat{A} \rangle=\beta$). 
The operation of DD using a photodetector is represented by the number operator, $\hat{N}$. Linearising the operator by assuming $\beta \gg 1$, we have
\begin{equation}\label{Eq:NumberOperator}
\begin{aligned}
\hat{{N}} & =\hat{{A}}^{\dagger}(t) \hat{{A}}(t) \\
& =(\beta+\delta \hat{{A}}^{\dagger}(t))(\beta+\delta \hat{{A}}(t)) \\
& =\beta^2+\beta \delta \hat{{A}}^{\dagger}(t)+\beta \delta \hat{{A}}(t)+\delta \hat{{A}}^{\dagger}(t) \delta \hat{{A}}(t) \\
& =\beta^2+\beta \delta \hat{{X}}(t)
\end{aligned}
\end{equation}
where $\delta \hat{X}(t) = \delta \hat{A}^\dagger (t) + \delta \hat{A}(t)$ is the $X$ quadrature operator, and we have neglected the higher-than-first-order terms in the modulations due to the assumption that the fluctuations are small (i.e., the contribution to the expectation value from the higher order terms is negligible). 
The last line of Eq.~(\ref{Eq:NumberOperator}) indicates that the use of DD effectively measures the $X$ quadrature. Eq.~(\ref{Eq:NumberOperator}) also explains the counter-intuitive concept of using DD to measure the $X$ quadrature -- in the same beam, the small signals (i.e., the modulation-induced fluctuations) beat with the carrier (i.e., the strong coherent component) and can be read out at a particular sideband frequency $\omega$. The technique for side-band readout proceeds as it would for homodyne detection and is described in, e.g.,~\cite{BRE97,RAL08}.

According to our Classical-Quantum Dual Encoding protocol, we are interested in the modulations around side-band frequency $\omega$, hence we focus our attention on the evolution of the Fourier space operator $\hat X_{\bar A}'(\omega)$ which is the Fourier transform of $\delta \hat{X}_{\bar A}'(t)$, representing the modulations of the field leaving Alice's station. We can write this Fourier space operator as:
\begin{align}
    \hat{X}_{\bar A}' &= X_C + \hat{X}_{\bar A},
    \label{XAD}
\end{align}
where $\langle X_C \rangle = \pm 2 \alpha$ is our discrete classical modulation, and $\hat{X}_{\bar A}$ represents quantum fluctuations.
After traversing the channel to Bob this operator evolves to
\begin{align}
    \hat{X}_B' &= \sqrt{\eta} \hat{X}_{\bar A}' + \sqrt{1- \eta} \hat{X}_v \nonumber \\
    & = \sqrt{\eta}(X_C + \hat{X}_{\bar A}) + \sqrt{1- \eta} \hat{X}_v ,
\end{align}
where $\eta$ is the transmission of the channel and $\hat{X}_v$ represents vacuum fluctuations that couple in due to the loss.
As will be described shortly it is arranged that the variance of the quantum fluctuations of the beam leaving Alice's station are at the QNL, i.e. $\langle \hat{X}_{\bar A}^2 \rangle = V_{\bar A} = 1$. As a result the signal to noise ratio (i.e. the signal power divided by the noise variance) of the classical signal received by Bob will be $SNR = \eta \langle X_C^2 \rangle/(\eta \langle \hat{X}_{\bar A}^2 \rangle + (1- \eta) \langle \hat{X}_v^2 \rangle) = 4 \eta  \alpha^2$, where we have used that the vacuum fluctuations also have unit variance. By choosing $\alpha$ sufficiently large the resulting bit error rate will be small and faithful transmission of the classical signals can be achieved.



\subsection{Quantum part}

To generate a key, we need to extract the quantum fluctuations from the combined classical and quantum signal that has been mixed down at our chosen side-band frequency $\omega$. Notice that the carrier power detected by Bob over the signal length will be proportional to $\eta \beta^2$ where $\eta$ is the channel transmission for that particular time-bin (assumed constant over the signal length). Thus $\eta$ can be measured shot-by-shot. Also, provided $\alpha$ is sufficiently large, we know $X_C$ shot-by-shot from the sign of the signal displacement. Thus we can extract the quantum signal by subtracting the weighted classical signal via
\begin{align}
    \hat{X}_B &= \sqrt{\eta} \hat{X}_{\bar A}' + \sqrt{1- \eta} \hat{X}_v - \sqrt{\eta} X_C\nonumber \\
    & = \sqrt{\eta} \hat{X}_{\bar A} + \sqrt{1- \eta} \hat{X}_v,
\end{align}
where we have used Eq.\ref{XAD} in going from the first to the second line. Note that the shot-by-shot carrier power also enables us to calibrate the QNL level for each signal.
The time bin length should be chosen appropriately; short enough so that the fluctuating power due to the atmosphere is approximately constant over the time bin but long enough to acquire sufficient signal strength.

\subsection{Global average quantum state}

We now analyse the generation and evolution of the quantum signal as shown diagrammatically in~\cref{fig:set-up}. As shown, Alice begins with two phase-locked beams, one of which has super-Poissonian amplitude noise, $\langle \hat{X}_a^2 \rangle = V_a > 1$, and the other has sub-Poissonian amplitude noise, $\langle \hat{X}_s^2 \rangle = V_s < 1$. One of the beams has a bright carrier -- it does not matter which but we will assume here that it is the super-Poissonian beam. The sub-Poissonian, or squeezed beam is intrinsically quantum, however the noise on the super-Poissonian beam could either be quantum, e.g. anti-squeezing or amplified spontaneous emission, or it could be induced via classical modulations. The most important thing is that the noise has a Gaussian distribution. Alice mixes her two beams on a beam-splitter, retaining the reflected beam which she directly measures and sending the transmitted beam (after imposing the classical signal) through the channel to Bob. Eve intercepts all the lost light. We can then write the evolution of the quantum parts of the amplitude quadrature operators as
\begin{align}
    \hat X_{A} &= \sqrt{1-\varepsilon} \hat X_a+\sqrt{\varepsilon}\hat X_s \\
    \hat X_E &= \sqrt{1-\eta} (\sqrt{\varepsilon} \hat X_a + \sqrt{1-\varepsilon} \hat X_s ) + \sqrt{\eta} \hat X_v \\
    \hat X_B &= \sqrt{\eta} (\sqrt{\varepsilon} \hat X_a + \sqrt{1-\varepsilon} \hat X_s ) + \sqrt{1-\eta} \hat X_v,
\end{align}
where $\varepsilon$ is the transmissivity of Alice's beamsplitter, $\eta$ is the transmissivity of the channel, $\hat X_{A}$, $\hat X_B$, and $\hat X_E$ are the $X$-quadrature operators for Alice, Bob, and Eve, respectively, and $\hat X_s$ and $\hat X_a$ are the $X$-quadrature operators for the squeezed (sub-Poissonian) and antisqueezed (super-Poissonian) beams, respectively.

For zero information leakage we require that the modulation Alice prepares to send to Bob has the same variance as the vacuum state (i.e. the QNL). That is, we require
\begin{align}
    V_{\bar A} &= \varepsilon V_a+(1-\varepsilon) V_s = 1.
\end{align}

For the modulation to be QNL we need
\begin{align}
    \varepsilon &= \frac{1-V_s}{V_a-V_s},\; 1-\varepsilon = \frac{V_a-1}{V_a-V_s}.
    \label{eq:no info}
\end{align}
Then,
\begin{align}
    \langle \hat X_A \hat X_B \rangle &= \sqrt{\eta}\sqrt{V_a-1}\sqrt{1-V_s}\\
    \langle \hat X_E \hat X_B \rangle &= 0.
\end{align}
That is, there are correlations in the $q$ quadrature between Alice and Bob, but there are no correlations in the $q$ quadrature between Eve and Bob. This means the Holevo information $\chi_{EB} = 0$. The Holevo information, $\chi_{EB}$, represents the maximum information that Eve can extract from her signal about Bob's measurements.
Since $\langle \hat X_E \hat X_B \rangle = 0$, Eve's system is not correlated with Bob's system in the $X$ quadrature.


Given no information leakage to Eve about Bob's measurement results we will find we only need Alice and Bob's covariance matrix. 
Alice and Bob's covariance matrix is as follows:
\begin{align}
 \Gamma_{AB} &= \left[ \begin{matrix}   \langle \hat X_A^2 \rangle    &    \langle \hat X_A \hat P_A \rangle    & \langle \hat X_A \hat X_B \rangle    & \langle \hat X_A \hat P_B \rangle   \\
\langle \hat P_A \hat X_A \rangle    &    \langle \hat P_A^2 \rangle    & \langle \hat P_A \hat X_B \rangle    &     \langle \hat P_A \hat P_B \rangle     \\
 \langle \hat X_B \hat X_A \rangle    &     \langle \hat X_B \hat P_A \rangle    &  \langle \hat X_B^2 \rangle   &  \langle \hat X_B \hat P_B \rangle     \\
   \langle \hat P_B \hat X_A \rangle    &  \langle \hat P_B \hat P_A \rangle   &  \langle \hat P_B \hat X_B \rangle    &  \langle \hat P_B^2 \rangle       \end{matrix} \right] \nonumber \\
   &= \left[ \begin{matrix}   a_q    &    0    & c_q    & 0   \\
   0    &    a_p    & 0    &     c_p     \\
  c_q    &    0    & b_q   & 0     \\
   0    & c_p   & 0    & b_p       \end{matrix} \right],\label{eq:CM_final}
\end{align}
where
\begin{align}
    a_q &= V_s\varepsilon + V_a(1{-}\varepsilon)\\
    a_p &= \frac{\varepsilon}{V_s} + \frac{1{-}\varepsilon}{V_a}\\
    b_q &= \eta_f[V_a\varepsilon + V_s(1{-}\varepsilon)] - \eta_f + 1 \nonumber \\
    & + \text{Var}(\sqrt{\eta})(V_a\varepsilon + V_s(1{-}\varepsilon) -1) \\
    b_p &= \eta_f\left( \frac{\varepsilon}{V_a} + \frac{1{-}\varepsilon}{V_s}\right)  - \eta_f+ 1 \nonumber \\ & + \text{Var}(\sqrt{\eta})\left( \frac{\varepsilon}{V_a} + \frac{1{-}\varepsilon}{V_s} - 1 \right) \\
    c_q &= \sqrt{\eta_f}\sqrt{\varepsilon}\sqrt{1{-}\varepsilon}(V_a{-} V_s )\\
    c_p &= \sqrt{\eta_f}\sqrt{\varepsilon}\sqrt{1{-}\varepsilon}\left( \frac{1}{V_a} {-} \frac{1}{V_s}\right). 
\end{align}
Here we have allowed for a fluctuating channel transmission as expected for an atmospheric channel, where $\eta_f = \langle \sqrt{\eta} \rangle ^2$ and Var$(\sqrt{\eta}) = \langle {\eta} \rangle - \eta_f$, and where $\langle \eta \rangle = \int_0^{\eta_\text{max}} \eta p(\eta) \; \text{d}\eta $ and $\langle \sqrt{\eta} \rangle = \int_0^{\eta_\text{max}} \sqrt{\eta} p(\eta) \; \text{d}\eta $. Notice that given the no information leakage condition (Eq.\ref{eq:no info}) we find $b_q = 1$, so the fluctuating channel transmission does not add noise to the quantum amplitude correlations.



\subsection{Asymptotic Secret key rate}\label{sec:asymptotic secret_key}

The secret key protocol we propose is an adaption of CV-QKD with reverse reconciliation \cite{GRO03}, using the no leakage protocol \cite{Jacobsen2018} and a passive eavesdropper \cite{PhysRevA.105.032602}. In the reverse reconciliation protocol Alice guesses the value of Bob's quadrature measurement outcomes based on her own outcomes. They then perform a reconciliation protocol which corrects errors in Alice's guesses at the cost of shortening the data string. In the limit of an infinitely long string this process does not leak any information to the eavesdropper -- this is called the asymptotic limit. The asymptotic secret key rate with reverse reconciliation is~\cite{devetak2005distillation}
\begin{align}
    K_{\text{asym.}} &= \beta I_{AB} - \chi_{EB},
\end{align}
where $\beta$ is the reverse reconciliation efficiency, $I_{AB}$ is Alice and Bob's mutual information, and $\chi_{EB}$ is a bound on Eve's maximal information given by the Holevo Bound. For the no leakage protocol with a passive eavesdropper we have $\chi_{EB}=0$ and so $K_{\text{asym.}} = \beta I_{AB}$.

The best possible performance of our key protocol will be in this asymptotic limit, with ideal reconciliation ($\beta=1$), strong squeezing ($V_s << 1$) and ideal detectors. Under these conditions we have $I_{AB} = \frac{1}{2}\log_2\frac{a_q}{a_q-{c_q^2}/{b_q}}$. Substituting the values in this limit we find the highest achievable rate of the zero-leakage protocol is
\begin{align}
    K_{\text{ideal}} &= -\frac{1}{2}\log_2{(1-\eta_f)}.
    \label{asym}
\end{align}
In order to include the effects of finite data block lengths a more sophisticated analysis is required, as will be introduced in Section \ref{sec:secret_key}.



\section{Free-space optical propagation and atmospheric effects}\label{Section:SimulationSettings}

 
In this section, we model realistic satellite-to-Earth channels via numerical simulations. In the following section we will use these results and realistic efficiencies and data block lengths to verify the feasibility and evaluate the performance of our proposed scheme. 

\subsection{System Model}\label{sec:system_model}
\begin{figure}
    \centering
    \includegraphics[width=1\linewidth]{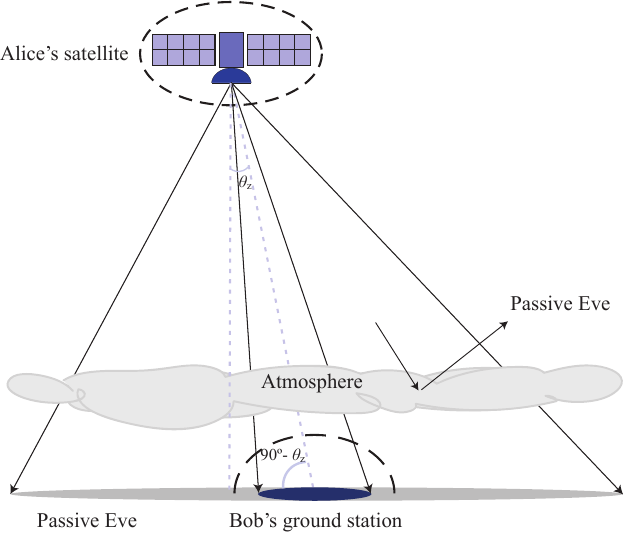}
    \caption{Satellite-to-Earth QKD link with a passive Eve. Classical and quantum information is encoded into a beam of light sent from the satellite (Alice) to the ground station (Bob). The channel is modelled as a free-space optical channel with the satellite at zenith angle $\theta_{\text{z}}$ (deg), as shown. 
    We model this downlink scenario as a concrete example, but our system is suitable for uplink or inter-satellite communications as well.}\label{fig:satellite}
\end{figure}

As a specific example, we consider the satellite-to-Earth communication setup shown in~\cref{fig:satellite} -- it should be noted that our system is not restricted to such an arrangement. In this work, the ground-station altitude is denoted as $h_0$, the satellite zenith angle at the ground station is denoted as $\theta_{\text{z}}$, and the satellite altitude at $\theta_{\text{z}}=0$ is denoted as $H$. The channel distance can then be given by $L = (H-h_0)/\cos\theta_{\text{z}}$. 
We assume a restricted eavesdropping scenario where the channel is monitored~\cite{https://doi.org/10.48550/arxiv.2212.04807}, and hence Eve is passive and not able to perform an active attack~\cite{e21040387}. However, as indicated in~\cref{fig:satellite}, we allow Eve to capture all the lost light.

The random inhomogeneities (i.e., turbulent eddies) within the Earth's turbulent atmosphere (which can be seen as a random medium) give rise to small random refractive index fluctuations (i.e., optical turbulence) that cause continuous phase modulations on an optical beam, imposing amplitude and phase distortions on the optical beam as it propagates through an atmospheric channel.
Specifically, the family of turbulent eddies whose sizes are bounded above by the outer scale $L_0$ and below by the inner scale $l_0$ constitutes the inertial subrange~\cite{book}.
It is commonly assumed that atmospheric turbulence satisfies 
\begin{equation}\label{Eq.OpticalTurbulenceAssumptions}
\langle n(\mathbf{R}) \rangle\!=\!1, \,\, \delta n(\mathbf{R})\!\ll\!1, \,\, \ensavg{\delta n(\mathbf{R})}=0,
\end{equation}
where $\mathbf{R}=[x,y,z]^{T}$ is the three-dimensional position vector, $n(\mathbf{R})$ denotes the the refractive index at $\mathbf{R}$, and $\delta n(\mathbf{R})\!=\!n(\mathbf{R}) - \langle n(\mathbf{R})\rangle$ denotes the small refractive index fluctuations~\cite{book}. 
Under the paraxial approximation and the assumptions in Eq.~(\ref{Eq.OpticalTurbulenceAssumptions}), the propagation of a monochromatic optical beam $\psi(\mathbf{R})$ through the turbulent atmosphere is governed by the stochastic parabolic equation~\cite{book}
\begin{equation}\label{Eq.sHe}
	\nabla_{\text{T}}^{2} \psi(\mathbf{R})+i 2 k \frac{\partial\psi(\mathbf{R})}{\partial {z}} +2 \delta n(\mathbf{R}) k^{2} \psi(\mathbf{R})=0,
\end{equation}
where $\nabla_{\text{T}}^{2}=\partial^{2} / \partial x^{2}+\partial^{2} / \partial y^{2}$ is the transverse Laplacian operator. 


In this work, we assume that all the transmitted optical beams are in the fundamental Gaussian mode. The general form of a Gaussian beam propagating along the $z$ axis is give by
\begin{equation}\label{Eq.Gaussian}
	\begin{aligned} \psi_{\text{G}}(\mathbf{r},z)&=\sqrt{\frac{2}{\pi}} \frac { 1 } { w ( z ) }  \exp \left[ \frac { - |\mathbf{r}| ^ { 2 } } { w ^ { 2 } ( z ) } \right]\exp \left[i\phi(\mathbf{r},z)\right],
	\end{aligned}
\end{equation}
where $\mathbf{r}=[x,y]^T$ denotes the two-dimensional position vector in the transverse plane, $w(z)=w_{0} \sqrt{1+(z / z_{\text{R}})^{2}}$, $w_0$ is the beam-waist radius, $z_{\mathrm{R}}=\pi w_{0}^{2}/\lambda$ is the Rayleigh range, $\lambda$ is the (central) optical wavelength, and $\phi(\mathbf{r},z) =  \frac {  k |\mathbf{r}| ^ { 2 } z } { 2 \left( z ^ { 2 } + z _ { \text{R} } ^ { 2 } \right) }  - \arctan \left( \frac { z } { z _ { \text{R} } } \right)$ with $k=2\pi/\lambda$ being the optical wavenumber. 

Although our scheme is, in principle, robust against the random phase distortions (recall that no local oscillator is required in our direct detection setup -- the wavefront matching between the local oscillator and the signal beam is not required), turbulence-induced effects related to amplitude distortions can still negatively affect the performance of our scheme. Indeed, these effects (e.g., beam wandering, beam-shape deformation, beam broadening, and intensity scintillation) can introduce substantial losses in real-world scenarios where a finite-sized aperture is used at the receiver~\cite{book}. Due to the random nature of atmospheric turbulence, such losses are also randomly fluctuating, giving rise to non-Gaussian excess noise and thus further degrading the system performance (see discussions in e.g.,~\cite{FadingChannelCV2012,FadingChannelCV2019}). 

We adopt the standard notation ``loss'' for visualization purposes; however, in our calculations, we always use the transmissivity $\eta$ (with the relation $\text{Loss}\,\text{[dB]} = -10 \log_{10} \eta$) to maintain consistency with our analyses in the previous sections. Assuming the transmitted beam is denoted by $\psi_{\text{G}}(\mathbf{r},0)$  and the received beam is denoted by $\psi_{\text{Rx}}(\mathbf{r},L)$, the (fluctuating) transmissivity is given by 
\begin{equation}\label{Eq.eta}
	\eta = \int_{\mathcal{A}} \left|\psi_{\text{Rx}}(\mathbf{r},L)\right |^2\,\text{d}^2\mathbf{r},
\end{equation}
where $L$ denotes the channel distance, and $\mathcal{A}$ denotes the aperture area at the receiver. Note that we assume a circular receiver aperture whose radius is $r_{\text{a}}$ in this work.

\subsection{Satellite-to-Earth atmospheric channel}
\subsubsection{Turbulence characterization}
The optical turbulence within the Earth's atmosphere must first be appropriately characterized. In this work, we adopt the widely used Hufnagel-Valley (HV) model to determine the structure parameter $C_n^2(h)$ that describes the strength of the optical turbulence within a satellite-based atmospheric channel. The HV model expresses $C_n^2(h)$ as the following function of altitude $h$~\cite{book}.
\begin{equation}\label{Eq.HV}
	\begin{aligned}
		C_{n}^{2}(h)& = 0.00594(v_{\text{rms}}/27)^{2}(h \times 10^{-5})^{10} \exp{(-h/1000)}\\
		&+2.7\!\times\!10^{-16} \exp{(-h/1500)}\!+\!A \exp{(-h/100)},
	\end{aligned}
\end{equation}
where $A$ is the ground-level turbulence strength in $\text{m}^{-2 / 3}$, and $v_{\text{rms}}$ is the root-mean-square {{(rms)}} wind speed in m/s. The rms wind speed is given by
\begin{equation}\label{Eq.Wind_Profile}
	v_{\text{rms}}=\left[\frac{1}{15 \times 10^{3}} \int_{5 \times 10^{3}}^{20 \times 10^{3}} V^{2}(h) \,\text{d}h\right]^{1 / 2},
\end{equation}
where $V(h)$ is the altitude-dependent wind speed profile. In this paper we adopt the Bufton wind profile~\cite{book}
\begin{equation}\label{Eq.Bufton}
	V(h)=V_{\text{g}}+30 \exp \left[-\left(\frac{h-9400}{4800}\right)^{2}\right],
\end{equation}
where $V_{\text{g}}$ is the ground-level wind speed.

Utilizing the structure parameter profile of the Earth's atmosphere, we adopt the scintillation index $\sigma_{\text{I}}^2$ (which is the normalized variance of intensity) and the Fried parameter $r_0$ (which quantifies the coherence length of turbulence-induced phase errors in the transverse plane) to quantify the effect of atmospheric turbulence on a propagating beam. For satellite-to-Earth channels, the scintillation index is given by~\cite{book}
\begin{equation}\label{Eq.ScintIdxTot}
	\sigma_{\text{I}}^{2}=\exp\!\left[\frac{0.49 \sigma_{\text{R}}^{2}}{\left(1+1.11 \sigma_{\text{R}}^{12 / 5}\right)^{7 / 6}}\!+\!\frac{0.51 \sigma_{\text{R}}^{2}}{\left(1+0.69 \sigma_{\text{R}}^{12 / 5}\right)^{5 / 6}}\right]-1,
\end{equation}
with $\sigma_{\text{R}}^2$ being the Rytov variance,
\begin{equation}\label{Eq.RytovVar}
	\sigma_{\text{R}}^{2}=2.25 k^{7 / 6} \sec ^{11 / 6}(\theta_{\text{z}}) \int_{h_{0}}^{H} C_{n}^{2}(h)\left(h-h_{0}\right)^{5 / 6} \,\text{d} h.
\end{equation}
For satellite-to-Earth channels, the Fried parameter is given by~\cite{book}
\begin{equation}\label{Eq.FriedParamTot}
	r_0=\left[0.423 k^{2} \sec \theta_{\text{z}} \int_{h_0}^{H} C_{n}^{2}(h) \,\text{d} h\right]^{-3 / 5}.
\end{equation}

The rate at which atmospheric turbulence changes with time is commonly quantified by the Greenwood frequency~\cite{Greenwood1977,CoherenceTimeZenith2014}
\begin{equation}
	f_{\text{G}}=2.31 \lambda^{-6 / 5}\left[\sec \theta_{\text{z}} \int C_n^2(h) V^{5 / 3}(h)\,\mathrm{d} h\right]^{3 / 5}.
\end{equation}
Correspondingly, the time interval over which atmospheric turbulence remains essentially unchanged can be quantified by the atmospheric coherence time~\cite{CoherenceTimeZenith2014}
\begin{equation}
	\tau_0 = \frac{0.134} {f_{\text{G}}}.
\end{equation}
 
\subsubsection{Method of numerical simulations}
In this work, the implementation of our numerical simulation largely follows our previous work~\cite{Ziqing_OAMQKD} (see,~e.g., the appendix of~\cite{Ziqing_OAMQKD} for more discussions). Specifically, we numerically solve Eq.~(\ref{Eq.sHe}) using the split-step method~\cite{SimAP} (also referred to as phase screen simulations) in order to faithfully capture the effects imposed by the optical turbulence on an optical beam propagating within the Earth's atmosphere. 
The main idea of this method is to model an atmospheric channel using a set of slabs with a random phase screen located in the midway of each slab. Two vacuum propagations with one random phase modulation in between are repeatedly performed until the beam reaches the receiver. 

The implementation of our phase screen simulations requires careful configuration and optimization based on the evaluation of the scintillation index (recall Eq.~(\ref{Eq.ScintIdxTot})) and the Fried parameter (recall Eq.~(\ref{Eq.FriedParamTot})) for both the whole channel and each individual slab.
In order to characterize the atmospheric turbulence within the $j^{\text{th}}$ slab, both the scintillation index (denoted as $\sigma_{\text{I}_j}^2$) and the Fried parameters (denoted as $r_{0_j}$) have to be evaluated locally for the turbulent atmosphere within that slab. 
For the actuate modelling of the atmospheric channel, we set the width of each slab based on the two conditions (i.e.,~$\sigma_{\text{I}_j}^2<0.1$ and $\sigma_{\text{I}_j}^2<0.1\sigma_\text{I}^2$) described in~\cite{Martin88}.

After determining the widths of the slabs, we use a well-known Fast-Fourier-Transform (FFT)-based spectral-domain algorithm~\cite{McGlamery67} to generate (random realizations of) the phase screen for each slab using its corresponding phase Power Spectral Density (PSD) function.
In this algorithm, the generation of each individual phase screen starts with the spectral-domain generation of a uniform square grid of random complex numbers sampled from a complex Gaussian distribution whose real and imaginary parts each have zero mean and equal variances with zero cross-covariances. 
This grid of random numbers is further weighted by the corresponding phase Power Spectral Density (PSD) function of the atmospheric turbulence within the corresponding slab. 
A FFT is then performed to transform this spectral-domain grid of weighted random complex numbers into a spatial-domain grid of random phase values (corresponding to an array of sample points of the phase screen) with the same statistics as the turbulence-induced phase variations within the corresponding slab.
In this work, we adopt the widely-used modified von Karman model for the the atmospheric turbulence, giving the following phase PSD function for the $j\,^{\text{th}}$ slab
\begin{equation}
    \Phi_{\phi_j}^{\mathrm{mvK}}(f)=0.023 r_{0_j}^{-5 / 3} \frac{\exp \left(-f^2 / f_{\text{m}}^2\right)}{\left(f^2+f_0^2\right)^{11 / 6}},
\end{equation}
where $f$ is the magnitude of the two-dimensional spatial frequency vector in the transverse plane in cycles/m, $f_0 =1/L_{\text{outer}}$, and $f_{\text{m}} = 0.9422/l_\text{inner}$~\cite{SimAP}.

For the free-space propagation, we utilize a physical optical propagation library named \textsc{proper}, whose routines implement FFT-based angular-spectrum and Fresnel-approximation methods to propagate a wavefront in near-field and far-field conditions, respectively~\cite{PROPER}. As a fully numerical approach to solving the stochastic parabolic equation, despite its computationally intensive nature, the split-step method can take into account most of the important effects (e.g.,~beam wandering, beam broadening, and scintillation) in atmospheric optical propagation. 
Although this method can also be adopted to characterize the temporal profile of the turbulence (i.e., the temporal turbulent fluctuations within one coherence time $\tau_0$), we notice that such a temporal profile characterization is not required within the scope of this work since an atmospheric channel can be considered constant for a large number of sequential signal pulses (more discussions will be provided in Sec.~\ref{sec:loss_statistics}).
The split-step method has been widely used to study the problem of atmospheric optical propagation under various channel conditions, providing quantitative agreement with both analytical results (e.g.,~\cite{OAM_Entanglement_MPS,EduardoEnhanced2021}) and real-world experimental data (e.g.,~\cite{EduardoDSTData2020}). 
More technical details regarding the implementation of our numerical simulations can be found in~\cite{Ziqing_OAMQKD,Ziqing_TMQKD,EduardoDSTData2020,EduardoEnhanced2021} and the references therein.




\subsection{Simulation settings}\label{SubSection:SimulationSettings}
We restrict ourselves to the case of a low-Earth-orbit (LEO) satellite with a satellite altitude $H=500\,\text{km}$, and we consider zenith angles $\theta_{\text{z}}$ ranging from $0^{\circ}$ to $60^{\circ}$. For the atmospheric parameters, we follow previous works (e.g.,~\cite{Ziqing_OAMQKD,Ziqing_TMQKD}) and set $A=9.6\times 10^{-14}\, \text{m}^{-2 / 3}$, $L_{\text{outer}} = 5\,\text{m}$, and $l_{\text{inner}}=1\,\text{cm}$. For the wind speed profile, we set $V_{\text{g}}=3\,\text{m/s}$, giving a value of $v_{\text{rms}}=21\,\text{m/s}$. For the optical parameters, we set $\lambda=1064\,\text{nm}$, and we set the beam-waist radius to $w_0=15\,\text{cm}$. 
Particularly, our setting of the $1064\,\text{nm}$ wavelength is due to the existence of high-quality sources of squeezed light at this wavelength (see, e.g.,~\cite{1064nmSqueezedLight}). This wavelength also falls within the atmospheric transmission window and is currently one of the most widely used operating wavelengths in real-world space-based optical communication systems (see comprehensive review papers, e.g.,~\cite{SpaceBasedFSO}).
For real-world implementations, we consider three reasonable receiver aperture radii, namely, $r_{\text{a}} = 15\,\text{cm}$, $r_{\text{a}} = 30\,\text{cm}$, and $r_{\text{a}} = 50\,\text{cm}$, at the ground station. We further fix the ground-station altitude to $h_0=0\,\text{m}$. In our numerical simulations, we generate 10000 independent realizations of the satellite-to-Earth channel under each simulation setting. Our simulation parameters are listed in Table~\ref{Tab:SimSettings}.
\begin{table*}[t]
\setlength{\tabcolsep}{6pt}
\renewcommand{\arraystretch}{1.2}
\caption{Simulation parameters.}\label{Tab:SimSettings}
\centering
\begin{tabular}{@{}ccccccccccc@{}}
\toprule
$A$ &
  $L_{\text{outer}}$ &
  $l_{\text{inner}}$ &
  $V_{\text{g}}$ &
  $v_{\text{rms}}$ &
  $\lambda$ &
  $w_0$ &
  $h_0$ &
  $H$ &
  $\theta_{\text{z}}$ &
  $r_{\text{a}}$ \\ \midrule
$9.6\times 10^{-14}\,\text{m}^{-2/3}$ &
  $5\,\text{m}$ &
  $1\,\text{cm}$ &
  $3\,\text{m/s}$ &
  $21\,\text{m/s}$ &
  $1064\,\text{nm}$ &
  $15\,\text{cm}$ &
  $0\,\text{m}$ &
  $500\,\text{km}$ &
  $0^{\circ}$ to $60^{\circ}$ &
  $\{15\,\text{cm},\,30\,\text{cm},\,50\,\text{cm}\}$ \\ 
  \bottomrule
\end{tabular}
\end{table*}

Although the existence of pointing errors can lead to an additional loss penalty in a real-world system, the resulting loss penalty can be significantly reduced by the use of state-of-the-art tracking techniques.
Indeed, via the use of feedback-based tracking techniques, the loss penalty resulting from pointing errors over a $1200\,\text{km}$ satellite-to-Earth channel is estimated to be smaller than $3\,\text{dB}$ (which is lower than the turbulence-induced loss and much lower than the diffraction loss) in a well-known real-world demonstration of satellite-to-Earth QKD~\cite{SatQKD}. Therefore, in this work we ignore the pointing-error-induced loss as we expect that including such a loss will not significantly change any of our observations.

\subsection{Channel loss statistics}\label{sec:loss_statistics}
Before presenting our main results, we first provide a visualization of the channel loss statistics (predicted from our numerical simulations) over a satellite-to-Earth channel. 

\begin{figure}
	\centering
	\includegraphics[width=0.9\linewidth]{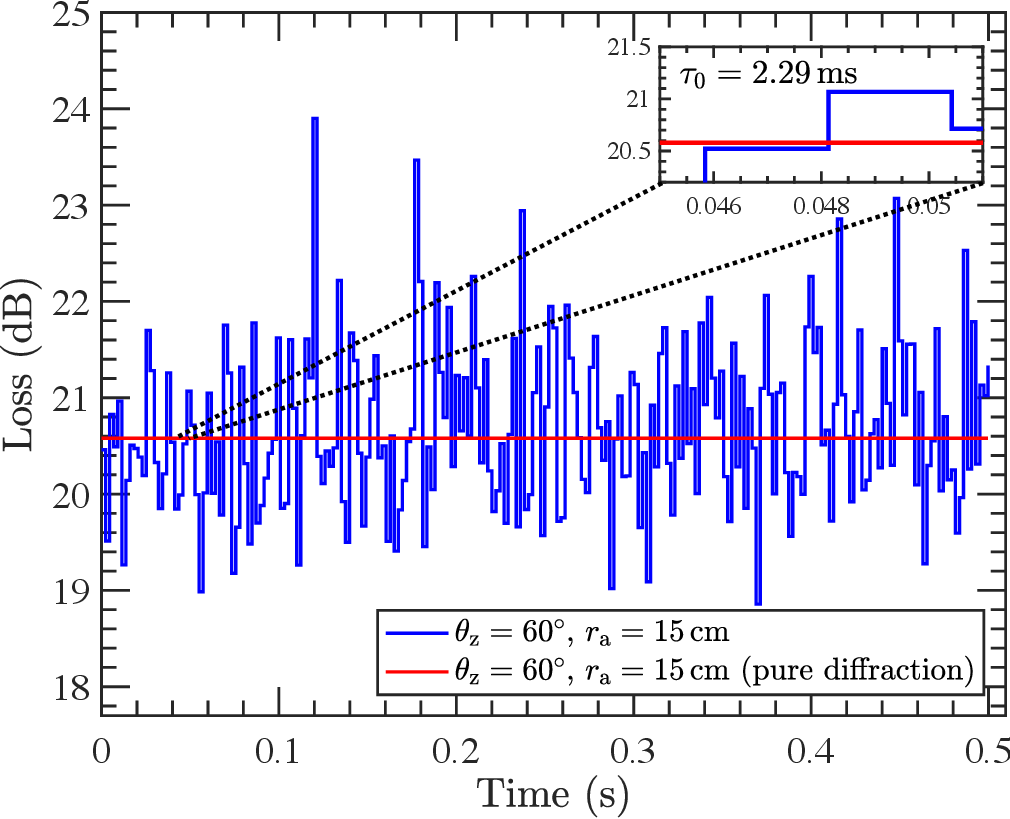}
	\caption{Fluctuating loss within a $0.5\,\text{s}$ time period over a satellite-to-Earth channel. The atmospheric coherence time is found to be $\tau_0=2.29\,\text{ms}$. Note that the channel loss is assumed to be constant within the atmospheric coherence time.}\label{fig:Fluctuations_Loss}
\end{figure}

To illustrate the time-dependent fluctuation of channel loss, in Fig.~\ref{fig:Fluctuations_Loss} we first plot the fluctuating loss over a satellite-to-Earth channel ($\theta_{\text{z}}=60^{\circ}$ and $r_{\text{a}}=15\,\text{cm}$) within a $0.5\,\text{s}$ time period.
It should be noted that Fig.~\ref{fig:Fluctuations_Loss} presents the channel loss fluctuation as a step function, with step size being the atmospheric coherence time $\tau_0$ -- this is because the channel loss can be considered constant within the time interval specified by $\tau_0$.
Specifically, the atmospheric coherence time under this setting is found to be $\tau_0=2.29\,\text{ms}$. Given that the coherence timescale of an atmospheric channel is on the order of milliseconds (see, e.g.,~\cite{book}) and that a pulsed source with a $100\,\text{MHz}$ repetition rate (which corresponds to a pulse period of $0.01\,\mu\text{s}$) is readily available, the channel can be considered constant for a large number ($\sim\!10^{5}$) of sequential signal pulses.
 
\begin{figure}
	\centering
	\includegraphics[width=0.9\linewidth]{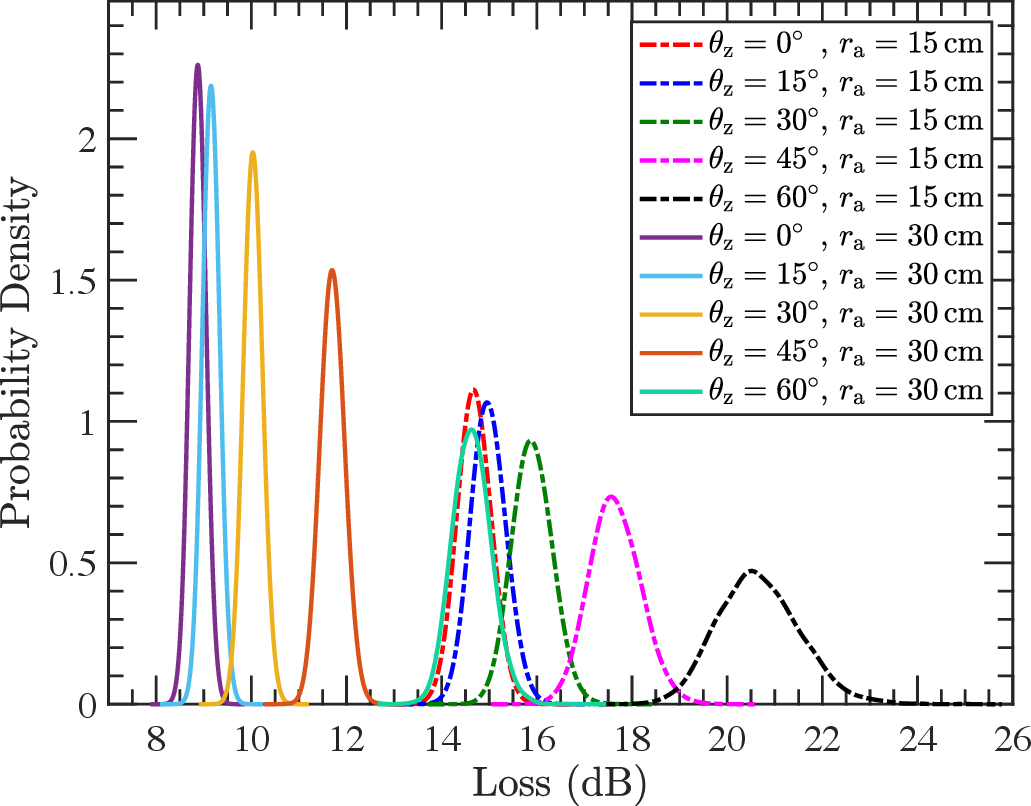}
	\caption{PDF of the channel loss achieved with different $r_\text{a}$ values under different $\theta_{\text{z}}$ values over a satellite-to-Earth channel. 
 The PDF curves are generated by fitting a kernel probability distribution to our discrete simulation data on channel loss using the \textit{fitdist} function provided by {\sc{Matlab}}.}\label{fig:PDF_Loss}
\end{figure}

In Fig.~\ref{fig:PDF_Loss}, we plot the Probability Density Function (PDF) of the channel loss, achieved with different $r_\text{a}$ values under different $\theta_{\text{z}}$ values, over a satellite-to-Earth channel.
From this figure, we can observe that both the average channel loss (indicated by the center of a PDF curve) and the fluctuation of channel loss (indicated by the spread of a PDF curve) decrease as the receiver aperture radius $r_{\text{a}}$ increases. 
The former observation is intuitive since a larger receiver aperture can capture a larger portion of the beam on the ground. 
The latter observation is due to the well-known effect of \textit{aperture averaging} in free-space optical communications -- the level of power fluctuation decreases as the receiver aperture size increases (more details can be found in many good textbooks, e.g.,~\cite{book}).
It can be also seen that under each setting, the distribution of channel loss is approximately centered at diffraction loss, and the spread of distribution (which indicates the fluctuation of channel loss) is generally minimal. 
Such observations indicate that diffraction loss (a deterministic loss) is the dominant source of loss over a satellite-to-Earth channel. Indeed, it is well known that beam wandering, which is the major source of turbulence-induced channel loss fluctuations, is negligible in a satellite-to-Earth channel (see, e.g.,~\cite{book,LEO_SatQ_2013,EduardoEnhanced2021,SatQKD}). 
This phenomenon is because an optical beam transmitted from a satellite becomes very large when it enters the Earth’s atmosphere, and the turbulence-induced effects are imposed on the optical beam only in the last segment of its propagation (note that the effective thickness of the atmosphere is only around $6\,\text{km}$ -- see, e.g.,~\cite{Atmosphere6km}).

\begin{figure}
	\centering
	\includegraphics[width=0.9\linewidth]{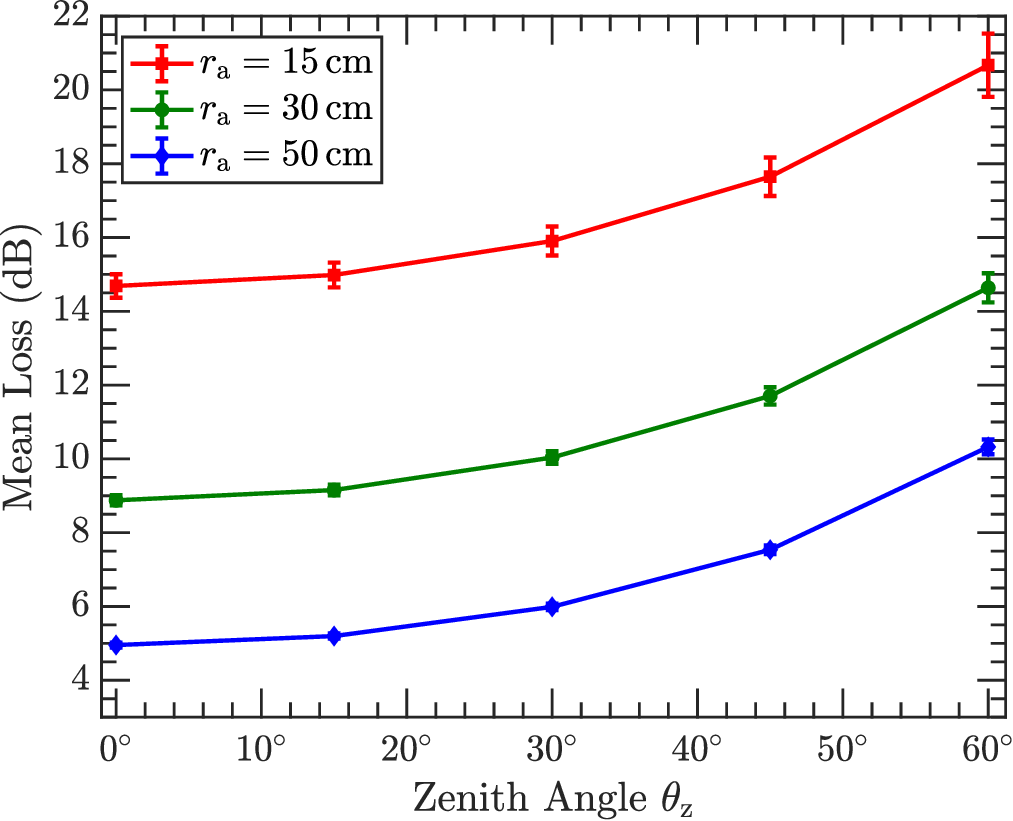}
	\caption{Mean channel loss achieved with different $r_{\text{a}}$ values under different $\theta_{\text{z}}$ values over a satellite-to-Earth channel. Error bars represent one standard deviation.}\label{fig:Loss_ZA}
\end{figure}

In Fig.~\ref{fig:Loss_ZA}, we plot the mean channel loss, achieved with different $r_\text{a}$ values under different $\theta_{\text{z}}$ values, over a satellite-to-Earth channel. The results in this figure are consistent with those in Fig.~\ref{fig:Fluctuations_Loss} and provide a quantitative representation of the mean channel loss and the channel loss fluctuation over a satellite-to-Earth channel.


\section{Secret key rate results}\label{sec:secret_key}

The composable, finite length secret key rate is reduced from the asymptotic rate due to finite size effects on the entropy, leaking of information through the public classical reconciliation of the raw key and the sacrifice of data bits in estimating the signal to noise (SNR) of the quantum signal. For our protocol it is given by ~\cite{PhysRevA.105.032602}
\begin{align}
    K &= {{1}\over{N}}(N' \beta I_{AB}-\sqrt{N'} \Delta_\text{AEP}-2\log_2[{1/(2\bar{\epsilon})}]),
\end{align}
which is $\epsilon$-secure~\cite{Leverrier_2015,PhysRevA.97.052327} where $\epsilon =2\epsilon_{\text{sm}} +\bar{\epsilon} + \epsilon_{\text{PE}}+ \epsilon_{\text{cor}}$. Here $\epsilon_{\text{sm}}$ is the entropy smoothing parameter and $\epsilon_{\text{cor}}$ and $ \epsilon_{\text{PE}}$ are the maximum failure probabilities for the error
correction and parameter estimation, respectively. For our protocol, Eve obtains no information from the optical communication hence parameter estimation is unnecessary and we can set $\epsilon_{\text{PE}} = 0$. In addition $\bar{\epsilon}$ is required by the left over Hash lemma.
As before, $\beta$ is the reverse reconciliation efficiency and $I_{AB}$ is Alice and Bob's mutual information. The block size is $N$ and $N'$ is the length of Bob’s string after the SNR estimation. Also, $\Delta_\text{AEP} = (d+1)^2+4(d+1)\sqrt{\log_2(2/ 2\epsilon_{\text{sm}}^2 )} +2 \log_2[2/( 2 \epsilon^2 \epsilon_{\text{sm}} )]+4 \epsilon_{\text{sm}}d/(\epsilon \sqrt{N'}  )$, where $d$ is the discretisation parameter used by Bob to extract key values. Alice and Bob's mutual information is 
\begin{align}
I_{AB} &= \frac{1}{2}\log_2\frac{a_q}{a_q-{c_q^2}/[{b_q+(1-\eta_B)v}]}, 
\end{align}
where $\eta_B$ is the efficiency of Bob's detector, and $v = 1+\frac{v_B}{1-\eta_B}$ with $v_B$ the electronic noise of the detectors.



\begin{figure}
    \centering
    \includegraphics[width=1\linewidth]{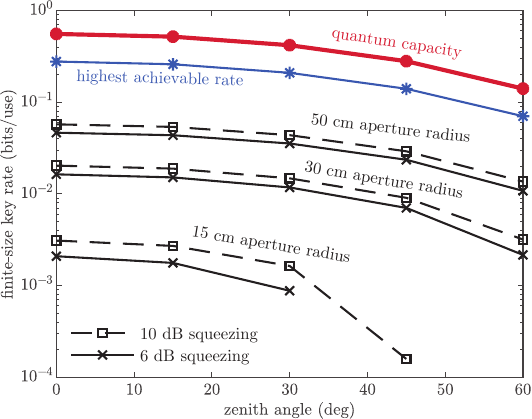}
    \caption{Finite-size secret key rate (bits/use) as a function of zenith angle (deg) for the squeezed-state protocol, secure against passive individual and collective attacks, with 6 dB and 10 dB of squeezing as shown. The beam-waist radius is set to 15 cm. The parameters are security parameter $\epsilon= 10^{-9}$, the discretisation parameter $d = 5$, Bob’s detector efficiency $\eta_B = 0.61$, electronic noise $\nu_B = 0.12$, reconciliation efficiency $\beta = 0.98$, 
    and block size $N = 10^{10}$, half of which is used for parameter estimation. The transmissivity data are simulated for 15 cm, 30 cm, and  50 cm receiver aperture radius size s, as outlined in~\cref{sec:loss_statistics}. The highest achievable rate of any QKD protocol (the quantum capacity or PLOB bound)~\cite{Pirandola_2017} and the highest achievable rate of our squeezed state protocol are also plotted (in red and blue, respectively, assuming a 50 cm aperture). Rates are simulated at discrete intervals only shown by points in the figure. Lines are added to guide the eye.}\label{fig:key_rate}
\end{figure}

The finite-size secret key rate is shown in~\cref{fig:key_rate}. We use state-of-the-art parameters for our calculations \cite{ZHA20} with the caveat that these parameters, in particular the reconciliation efficiency, may be difficult to realize in the field. The parameter are: security parameter $\epsilon= 10^{-9}$; the discretisation parameter $d = 5$; Bob’s detector efficiency $\eta_B = 0.61$; electronic noise $\nu_B = 0.12$; reconciliation efficiency $\beta = 0.98$; 
and block size $N = 10^{10}$, half of which is used for parameter estimation. These calculations use simulated transmissivity data (10000 realisations) as discussed in~\cref{sec:loss_statistics}. With the simulated data and parameters outlined in the caption of~\cref{fig:key_rate},  we obtain positive secret key rate for 30 and 50 cm receiver aperture radius sizes, and obtain positive key for 15 cm receiver aperture radius size but not for large zenith angles. Turbulence has little impact on the key rate. In~\cref{fig:key_rate}, we also plot the capacity (red) and the ideal asymptotic key rate (Eq.\ref{asym}) (blue) for comparison. The capacity (i.e. the ultimate limit given by the two-way (classically-assisted) quantum capacity of the pure-loss channel~\cite{Pirandola_2017}) and the asymptotic key rate are both plotted using loss parameters for the $50$ cm aperture radius.
We can estimate that for larger block sizes, say $N=10^{14}$ 
uplink becomes possible since the maximum amount of tolerable loss is then about 40 dB. 

\section{Conclusion}\label{sec:conclusion}

We have considered a typical optical classical communication scheme based on amplitude modulation and direct detection, and incorporated a simultaneous CV QKD protocol. With the assumption of a passive Eve, we predict positive finite size secret key rates for realistic downlink free-space channels without affecting the SNR of the classical data.
Given a classically monitored channel active eavesdropper attacks are deemed unrealistic, so the security of the secret key is strong. No optical local oscillator is required for our protocol.

Our scheme requires the injection of squeezed light at the sender and efficient, low noise photo-detection at the receiver in order to enable encoding and read out of a quantum signal with a QNL variance. The purest current sources of squeezing are inefficient and so could add significantly to the local energy budget of the satellite. However, as the squeezing is only needed on one quadrature, purity is not a major issue, and so in principle a high efficiency squeezed laser \cite{YAM92} might be used as the source (i.e. $\hat X_s$ in Fig.\ref{fig:set-up}), adding very little to the local energy budget.  

In principle, given a passive Eve, a super-QNL encoding variance is possible, avoiding the need for squeezing. However, there are the following significant advantages to using a QNL encoding: (i) greater range and higher key rate for the QKD; (ii) the classical signals always have a QNL noise floor; (iii) the atmospheric fluctuation noise is cancelled; and (iv) there is reduced computational overhead and increased efficiency in the reconciliation protocol due to the zero information leakage of raw key information to Eve.

We believe our system features the minimum addition of quantum resources needed to combine classical communication and QKD in a system with practical performance
and good compatibility with existing satellite-based communication infrastructure. As such we expect it may play a role in next generation LCS systems.\\

\section*{Acknowledgements}

The Australian Government supported this research through the Australian Research Council's Linkage Projects funding scheme (project LP200100601). The views expressed herein are those of the authors and are not necessarily those of the Australian Government or the Australian Research Council. Approved for Public Release; Distribution is Unlimited; \#23-2133.


\bibliography{bibfile.bib}

\end{document}